\documentclass[prd,aps,preprint,superscriptaddress,floatfix,preprintnumbers]{revtex4-1}
\usepackage{mathrsfs}
\usepackage{amsfonts}
\usepackage{amsmath,amssymb,bm} 
\usepackage{array}
\usepackage{verbatim}
\usepackage{epsfig}
\usepackage{latexsym}
\usepackage{graphicx,graphics,color} 
\usepackage{hyperref}
\hypersetup{colorlinks, linkcolor = [rgb]{0,0.0,0.5}, citecolor = [rgb]{0,0.0,0.5}, urlcolor = [rgb]{0,0.0,0.5}}
\usepackage{url}
\usepackage[normalem]{ulem}
\usepackage{xcolor}
\usepackage{ulem}
\usepackage{longtable}
\usepackage{multirow}
\usepackage{hhline,multirow,tabularx}
\usepackage{dcolumn}
\usepackage{slashed}
\usepackage{rotating}
\usepackage{mathrsfs}
\usepackage{xr}
\usepackage{physics}

\newcommand{\fref}[1]{Fig.~\ref{f.#1}}

\newcommand{\eref}[1]{Eq.~(\ref{e.#1})}

\begin{document}

\preprint{JLAB-THY-20-3230}

\title{\mbox{Machine learning-based event generator for electron-proton scattering}}

\author{Y.~Alanazi}
\affiliation{Department of Computer Science, Old Dominion University, Norfolk, Virginia 23529, USA}
\author{P.~Ambrozewicz}
\affiliation{Jefferson Lab, Newport News, Virginia 23606, USA}
\author{M.~Battaglieri}
\affiliation{Istituto Nazionale di Fisca Nucleare, 16146 Genova, Italy}
\author{A.~N.~Hiller~Blin}
\affiliation{Institute for Theoretical Physics, T\"{u}bingen University, Auf der Morgenstelle 14, 72076 T\"{u}bingen, Germany} 
\author{M.~P.~Kuchera}
\affiliation{\mbox{Department of Physics, Davidson College, Davidson, North Carolina 28035, USA}}
\author{Y.~Li}
\affiliation{Department of Computer Science, Old Dominion University, Norfolk, Virginia 23529, USA}
\author{T.~Liu}
\affiliation{{Key Laboratory of Particle Physics and Particle Irradiation (MOE), Institute of Frontier and Interdisciplinary Science, Shandong University, Qingdao, Shandong 266237, China}}
\author{R.~E.~McClellan}
\affiliation{Jefferson Lab, Newport News, Virginia 23606, USA}
\author{W.~Melnitchouk}
\affiliation{Jefferson Lab, Newport News, Virginia 23606, USA}
\author{E.~Pritchard}
\affiliation{\mbox{Department of Physics, Davidson College, Davidson, North Carolina 28035, USA}}
\author{M.~Robertson}
\affiliation{Department of Mathematics and Computer Science, Davidson College, Davidson, North Carolina 28035, USA}
\author{N.~Sato}
\affiliation{Jefferson Lab, Newport News, Virginia 23606, USA}
\author{R.~Strauss}
\affiliation{Department of Mathematics and Computer Science, Davidson College, Davidson, North Carolina 28035, USA}
\author{L.~Velasco}
\affiliation{Department of Physics, University of Dallas, Irving, Texas 75062, USA}

\begin{abstract}
We present a new machine learning-based Monte Carlo event generator using generative adversarial networks (GANs) that can be trained with calibrated detector simulations to construct a vertex-level event generator free of theoretical assumptions about femtometer scale physics. 
Our framework includes a GAN-based detector folding as a fast-surrogate model that mimics detector simulators. 
The framework is tested and validated on simulated inclusive deep-inelastic scattering data along with existing parametrizations for detector simulation, with uncertainty quantification based on a statistical bootstrapping technique.
Our results provide for the first time a realistic proof-of-concept to mitigate theory bias in inferring vertex-level event distributions needed to reconstruct physical observables.  
\end{abstract}

\date{\today}
\maketitle

\section{Introduction}

Since the early 1970s, Monte Carlo event generators (MCEGs) have played a vital role in facilitating studies of QCD in high-energy scattering processes.
From the experimental perspective, MCEGs are a crucial part of the procedure used for modeling the detector response folded into measured quantities (``detector-level'') to  extract the {\rm true} energies and momenta of final state particles as produced at the interaction point (``vertex-level'').
The development of modern MCEGs, such as PYTHIA~\cite{Sjostrand:2007gs}, HERWIG~\cite{Bahr:2008pv}, and SHERPA~\cite{Gleisberg:2008ta}, has been driven by a combination of high-precision experimental data and theoretical inputs.
The latter have involved a mix of perturbative QCD methods, describing the dynamics of quarks and gluons at short distances, and phenomenological models that map the transition from quarks and gluons to observable hadrons, as well as nonperturbative inputs such as parton distribution functions for applications involving hadrons in the initial state~\cite{Jimenez-Delgado:2013sma, Forte:2013wc, Ball:2017nwa, Hou:2019efy, Sato:2019yez, Bailey:2020ooq}. 
While the theoretical  assumptions are usually well justified, an approach that mixes data with a model for the underlying physical law which we want to infer can potentially lead to biased results.
Moreover, the need to correct for detector effects typically becomes increasingly difficult in higher dimensions and prevents a faithful reconstruction of vertex level events in a model independent way.
In this work we present a novel approach to build an event-level interpolation tool based on machine learning (ML) that avoids theoretical assumptions about the femtometer-scale physics and a strategy to correct for detector effects at the event level.

MCEGs in general can be viewed as a type of data compactification utility, encapsulating large amounts of data collected from multiple experiments which can be regenerated.
On the other hand, the reliance of existing MCEGs on theoretical assumptions of factorization and hadronization models limits their ability to capture the full range of possible correlations between the produced particles' momenta and spins.
Moreover, existing theory-based MCEGs are limited in the scope of applications.  
For instance, to date no MCEG is able to reproduce all the possible single-spin or double-spin asymmetries in inclusive or semi-inclusive electron-proton deep-inelastic scattering (DIS).

Having an MCEG that faithfully simulates particle reactions by preserving all the correlations among the particles' momenta is extremely valuable for theoretical developments. 
In practice, such correlations are integrated out or projected onto customized degrees of freedom, which effectively limits the ability to fully test the theory.
In processes such as semi-inclusive DIS, for example, different regions of phase space are expected to be dominated by different physical mechanisms, and having access to all correlations among the particles' momenta is essential to understanding these mechanisms.

In this paper we present a strategy for constructing an ML-based event generator (MLEG) using generative adversarial networks (GANs) \cite{Goodfellow:2014upx}, which have been increasingly utilized recently in high-energy physics applications as a tool for fast Monte Carlo simulations~\cite{Paganini_2018, CaloGAN_2018, de_Oliveira_2017, Musella_2018, hashemi2019lhc, butter2019GAN, MMD_2007}.
A detailed survey of MLEGs for physics event generation can be found in Ref.~\cite{MLEG_survey}. 
A crucial feature of GANs (as well as generative models in general) is their ability to generate synthetic data by learning from real samples without explicitly knowing the underlying physical laws of the original system.  
We present a case study for inclusive DIS with realistic pseudodata generated from phenomenological models.
We first train the MLEG that can faithfully reproduce the phase space of inclusive DIS along with uncertainty quantification (UQ) stemming from finite statistics and model architectures.
Subsequently, we implement detector effects using an effective parametrization of detectors and train the MLEG and folding algorithms to simulated detector-level DIS events.
For the first time a closure-test for reconstructing vertex-level DIS events, free of theoretical assumptions, is performed.
The results provide a new opportunity for experimental data analysis to use the GAN approach to build theory-free event generators which mitigate biases induced in reconstructing physical observables from experimental data. 
Moreover, the technique provides a new form of data representation that can be easily distributed, in contrast to the traditional data-representation via histograms that are limited for processes with high-dimensional phase space.

We begin the discussion in Sec.~\ref{secII} with a schematic overview of the MLEG training with our GAN-based event-level interpolator.
This is followed in Sec.~\ref{secIII} by a description of the ML detector surrogate that we use in order to simulate the effects of real particle detectors. 
The application to inclusive electron-proton DIS is discussed in Sec.~\ref{secIV}, where we examine GAN training both without and with detector effects.
In Sec.~\ref{secV} we summarize our findings and discuss future extensions and applications. \\ \\

\section{GAN-based event-level interpolator}
\label{secII}

\begin{figure}[t!] 
    \centering
    {\includegraphics[width=0.96\textwidth,trim={0 3cm 0 2cm}, clip]{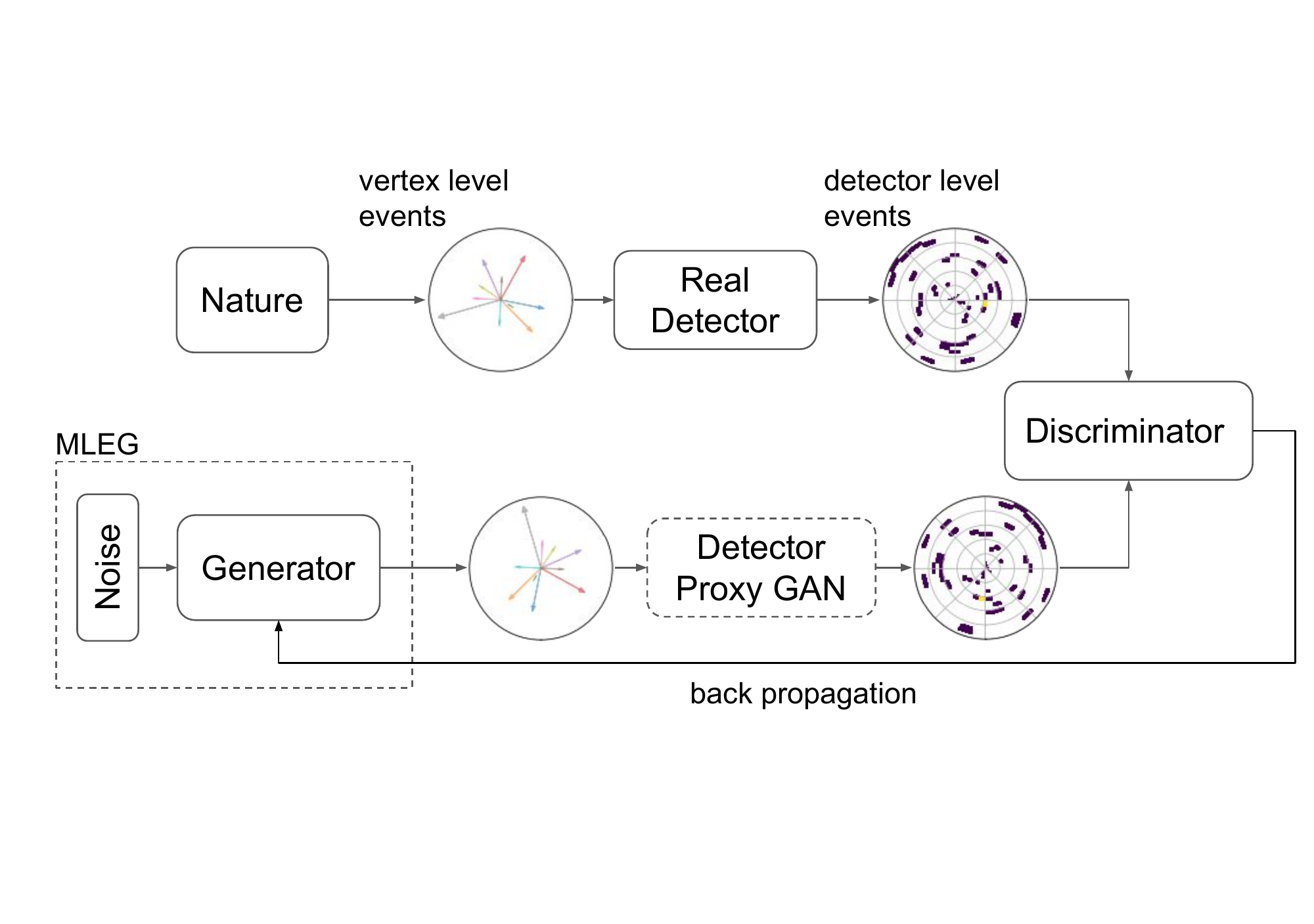}}
    \caption{Schematic view of the MLEG GAN training framework. The MLEG (dashed box) uses a generator which transforms noise into event-level features. The generator is concatenated with a detector simulator to mimic synthetic detector-level event features. The deep neural network based discriminator compares detector-level event features in order to build gradients to update the generator of the MLEG.} 
    \label{f.MLEG}
\end{figure}

A schematic view of the training workflow of our MLEG GAN is illustrated in \fref{MLEG}, where, as usual, the GAN model is composed of a {\it generator} and a {\it discriminator}. 
The generator converts noise through a deep neural network into event-level features, which is customized by a given reaction.
The generated event features are then passed into a detector simulator to convert them as ``trial'' detector-level events. 
The discriminator learns through another deep neural network to differentiate the true detector-level event samples from the ones produced by the generator and the detector simulator.
The GAN training evolves as the generator and discriminator compete adversarially, each updating their parameters during the training process. 
Eventually, the generator is able to produce synthetic samples that the discriminator can no longer distinguish from the real samples, at which point the training of the MLEG is complete.

Although GANs have demonstrated impressive results in various applications, including generating near-realistic images~\cite{karras2018stylebased}, music~\cite{mogren2016crnngan}, and videos~\cite{clark2019adversarial}, training a successful GAN model is known to be notoriously difficult. 
Many GAN models suffer from major problems, such as mode collapse, non-convergence, model parameter oscillation, destabilization, vanishing gradient, and over-fitting due to unbalanced training of the generator and discriminator. 
Approaches and techniques to address these general problems have been proposed and discussed recently in the literature~\cite{Salimans2016ImprovedTF, Arora2017DoGA, Arjovsky2017TowardsPM, srivastava2017veegan, bang2018mggan}.

Unlike common GAN applications, such as the generation of realistic high resolution images, the success of our GAN application as nuclear and high-energy physics event generators relies on its ability to faithfully reproduce correlations among the particles' momenta, which are increasingly difficult in higher (greater than one or two) dimensions.
At the same time, the corresponding multidimensional momentum distributions or histograms display rapid changes in the phase space that spans several orders of magnitude.
The challenge is then to design suitable GAN architectures capable of reproducing all of the correlations among the particles, along with a faithful reproduction of the multidimensional histograms across the phase space.
In Sec.~\ref{secIV} we will discuss in detail about how to customize this for our specific application of inclusive DIS.

\section{ML detector surrogate}
\label{secIII}

\begin{figure}[t] 
    \centering
    {\includegraphics[width=0.96\textwidth,trim={2cm 2cm 0 1cm}, clip]{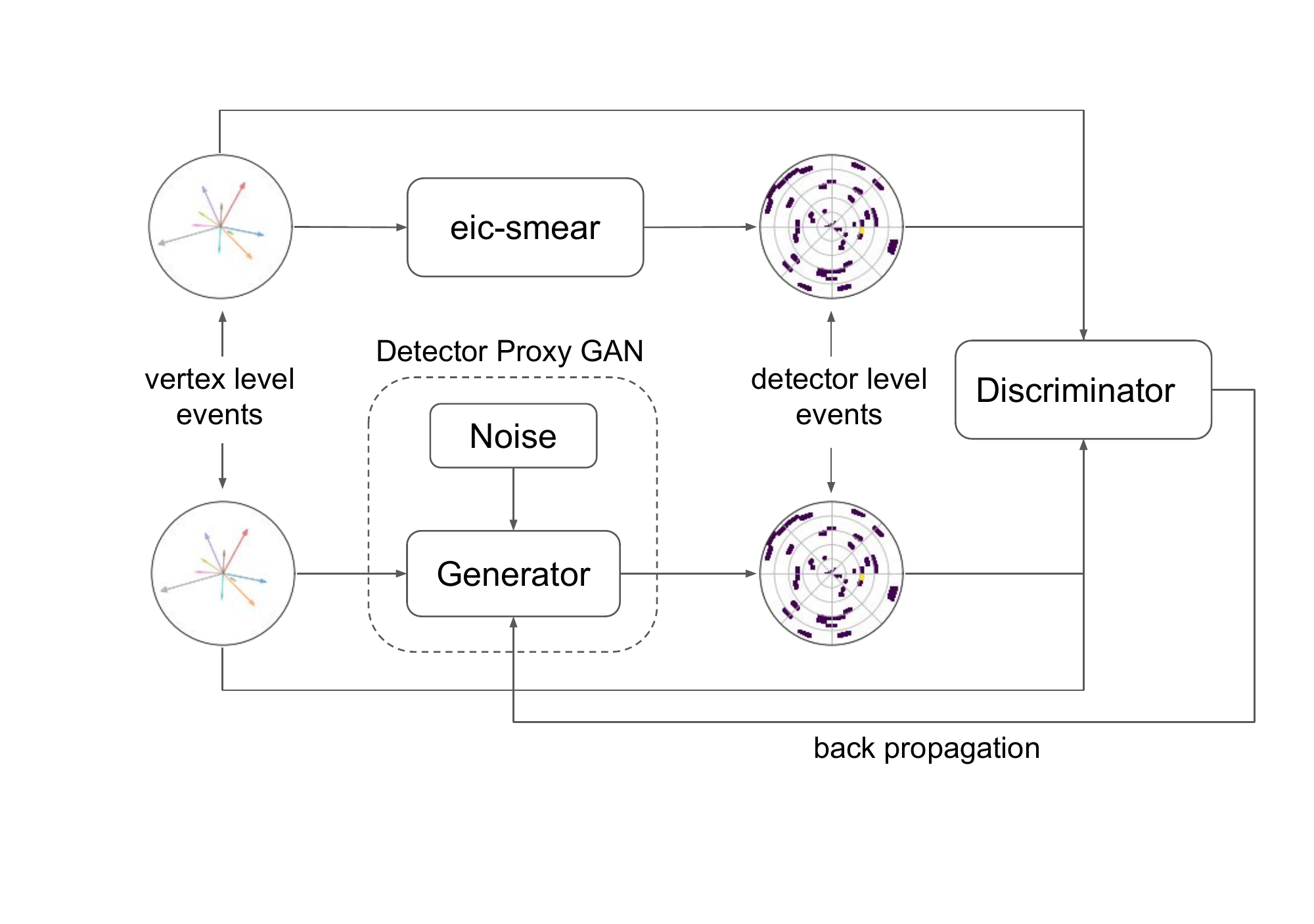}}
    \caption{Schematic view of the ML detector surrogate, where a generator converts input vertex-level event features and noise to detector-level event features. The training samples are obtained from guess vertex-level samples and the corresponding detector-level samples using a detector simulator. The discriminator (right hand side of the figure) is trained simultaneously with vertex-level and detector-level event features in order to minimize the dependence of the generator on the input vertex-level guess samples.}
    \label{f.folding}
\end{figure}

Experimental data, provided in the form of final state particle momenta, are affected by distortions introduced by experimental detectors.
A correction procedure is usually necessary to extract the {\rm true} information from the measured cross sections and provide the vertex-level distributions used in physics analysis.
Such detector effects have multiple causes, including limited acceptance, finite resolution, efficiency distortion, and bin migrations due to radiation and rescattering.
Corrections are commonly taken into account using unfolding procedures that attempt to correct for the detector effects at the histogram level, which  requiring {\it ad hoc} corrections for each type of observable.

In order to demonstrate that our framework is realizable in a real experimental analysis, such detector effects must be incorporated.
For this purpose, we use the {\it ``eic-smear''} software package \cite{eicsoft}, which was developed at Brookhaven National Laboratory as a fast simulation tool for the future Electron-Ion Collider \cite{EICyellowreport}, and provides a simplified parametrization of the response of the detectors.

We develop ML-based detector surrogates using a secondary conditional GAN, as illustrated in \fref{folding}.
The idea is to train a conditional generator simulating the smearing effect of the detector by converting input vertex-level event features and noise into detector-level event features, as dictated by {\it eic-smear}.
To do this we build training samples using trial vertex-level guess event samples and the associated {\it eic-smear} detector-level samples to train the conditional-GAN.
Once the conditional GAN is trained, the ML detector surrogate (represented by the dashed box in \fref{folding}) can be integrated as the detector simulator in \fref{MLEG}.  
It is worth noting that for a more realistic description of detector effects, the {\it eic-smear} parametrization should be replaced by a full GEANT-based~\cite{geant4} detector model.
However, its integration within our MLEG models using standard ML libraries is beyond the scope of the present analysis, and will be the subject of future work.

\section{Application to inclusive electron-proton scattering}
\label{secIV}

In this section we describe the application of our MLEG strategy to the inclusive unpolarized DIS of electrons (four-momentum $k$) from protons (four-momentum $P$).
Our goal is solely to produce the scattered electron phase space, labeled by the four-momentum $k'$.
As a surrogate for real experimental data, we use pseudodata generated from the JAM QCD global analysis framework \cite{Cocuzza:2021rfn} that has been tuned to describe world data on inclusive DIS and other high-energy scattering processes.

The inclusive electron DIS samples are generated at a center of mass energy of 318.2~GeV, compatible with HERA kinematics, by integrating the 2-dimensional differential cross section $\dd{\sigma}/\dd{x}\dd{Q^2}$, computed at next-to-leading order in perturbative QCD using importance sampling, and unweighting events over a very dense binning in $(x,Q^2)$-space.
Each event is transformed into an outgoing electron momentum in the HERA laboratory frame by generating an azimuthal angle relative to the beam axis sampled from a uniform distribution.
While our ultimate goal is to apply this approach to real data, this case study provides unique insights of our ML workflow and allows us to identify challenges in formulating a suitable feature space to be learned by the model.

When training the GAN solely using the electron momentum in the laboratory frame as event features, the generator was found to create electron samples that violate momentum conservation near the edge of the phase space, and the model was not sensitive enough to prevent the production of these samples \cite{MLEG_FAT-GAN}.
To alleviate this problem and aid the training, we use a change of variables that enhances the discriminator awareness in these difficult regions. 
Specifically, we define the scaled variables   
\begin{subequations}
\label{e.nu12}
\begin{align}
    \nu_1 &= \ln\big((k'_0-k'_z)/1\,{\rm GeV}\big),  \\
    \nu_2 &= \ln\big((2 E_e - k'_0 - k'_z)/1\,{\rm GeV}\big),
\end{align}
\end{subequations}
where $E_e$ is the incident electron energy, $k'_0$ and $k'_z$ are scattered electron energy and longitudinal momentum, respectively.
In Eqs.~(\ref{e.nu12}) the energies and momenta in the arguments of the log are implicitly in units of GeV.
These variables can be easily inverted into the original momentum space. 
In particular, the variable $\nu_2$ changes rapidly as the energy of the outgoing electron approaches its limit, allowing the discriminator to be aware of such region.

In the following, we present details of our chosen ML architecture used for the event-level interpolation and the ML detector surrogate.

\begin{itemize}

    \item \textbf{MLEG}: The input to the generator in \fref{MLEG} is a 100-dimensional white noise array centered at 0 with unit standard deviation.
    The generator network consists of 5 hidden dense layers, with 512 neurons per layer, activated by a leaky Rectified Linear Unit (ReLU) function. 
    The number of layers and neurons are optimized to balance execution time and convergence. 
    The last hidden layer is fully connected to a 2-neuron output corresponding to the variables $\nu_1$ and $\nu_2$, activated by a linear function representing the generated features. 
    The corresponding discriminator also consists of 5 hidden dense layers with 512 neurons per layer, optimized as for the generator, and activated by a leaky ReLU function. 
    To avoid overfitting, a 10\% dropout rate is applied to each hidden layer.
    The last hidden layer is fully connected to a single-neuron output, where ``1'' indicates a true event and ``0'' a fake event. 
    The discriminator $D$ is trained to give $D(\bm{F}) = 1$ for each training sample $\bm{F}$, and $D(\widetilde{\bm{F}}) = 0$ for each sample $\widetilde{\bm{F}}$ produced by the generator.

    \item \textbf{ML detector surrogate}: The detector surrogate model is based on a conditional GAN architecture~\cite{mirza2014conditional}.
    As shown in \fref{folding} we have a generator that receives vertex-level as input in addition to a 100-dimensional white noise centered at 0 with unit standard deviation. 
    The generator will learn to fold the inputs and produce detector-level events that mimic the detector response dictated by {\it eic-smear}.
    By conditioning the model on vertex-level event features we can enforce learning the correlations between vertex and detector level events as opposed to learning a deterministic mapping between inputs and outputs.
    As for the MLEG, the generator will produce a 2-neuron output corresponding to the detector-level variables $\nu_1$ and $\nu_2$, activated by a linear function representing the generated features, and the discriminator will similarly produce ``0'' or ``1'' for training and generated samples, respectively. 
    In both the generator and discriminator architectures of the ML detector surrogate, we use the same number of hidden layers, neurons, dropout rates, and activation functions as in our MLEG.
    A similar idea of using GAN for detector effects has been proposed by Bellagente~{\it et~al.}~\cite{butterfolding}, where in contrast to our folding procedure, parton-level data is mapped to detector-level data using a conditional GAN model.
    
\end{itemize}

For both of our GAN architectures we adopt the Least Squares GAN (LSGAN) \cite{mao2017squares}, which replaces the cross entropy loss function in the discriminator of a regular GAN by a least squares term, 
\begin{subequations}
\begin{align}
    \min_D V(D) 
    &= \frac12 \big\langle( D(x)-b)^2 \big\rangle_{x \sim P_T}
     + \frac12 \big\langle( D(G(\tilde{x}))-a)^2 \big\rangle_{\tilde{x} \sim P_G}\, ,  \\
    \min_G V(G) 
    &= \frac12 \big\langle(D(G(x))-c)^2\big\rangle_{x \sim P_G}\, ,
\end{align}
\end{subequations}
where $P_G$ denotes the distribution of the generated samples and $P_T$ the distribution of the training samples.
As a result, by setting $b-a=2$ and $b-c=1$, minimizing the loss function of LSGAN implies minimizing the Pearson $\chi^2$ divergence.
For the conditional model, the objective functions can be defined as
\begin{subequations}
\begin{align}
    \min_D V(D) 
    &= \frac12 
       \big\langle( D(x|y)-b)^2 \big\rangle_{x \sim P_T,\, y \sim P_v}\,
    +\,\frac12 
    \big\langle( D(G(\tilde{x}|y))-a)^2 \big\rangle_{\tilde{x} \sim P_G,\, y \sim P_v}\, ,  \\
    \min_G V(G) 
    &= \frac12 \big\langle(D(G(x|y))-c)^2\big\rangle_{x \sim P_G,\, y \sim P_v}\, ,
\end{align}
\end{subequations}
where $P_v$ denotes the conditioned vertex-level samples that are fed as inputs to the ML detector surrogate.
The main advantage of the LSGAN is that by penalizing the samples that are far from the decision boundary, the generator is prompted to generate samples closer to the manifold of the true samples.

Our networks are trained adversarially for 100,000 epochs, where an epoch is defined as one pass through the training data set. 
For the optimizer, in both cases we use Adam \cite{kingma2014adam} with a $10^{-4}$ learning rate, $\beta_1 = 0.5$, and $\beta_2 = 0.9$. 
To balance the generator and discriminator training, the training ratio is set to~5.

\subsection{GAN training without detector effects}

\begin{figure}[t] 
    \includegraphics[width=0.8\textwidth]{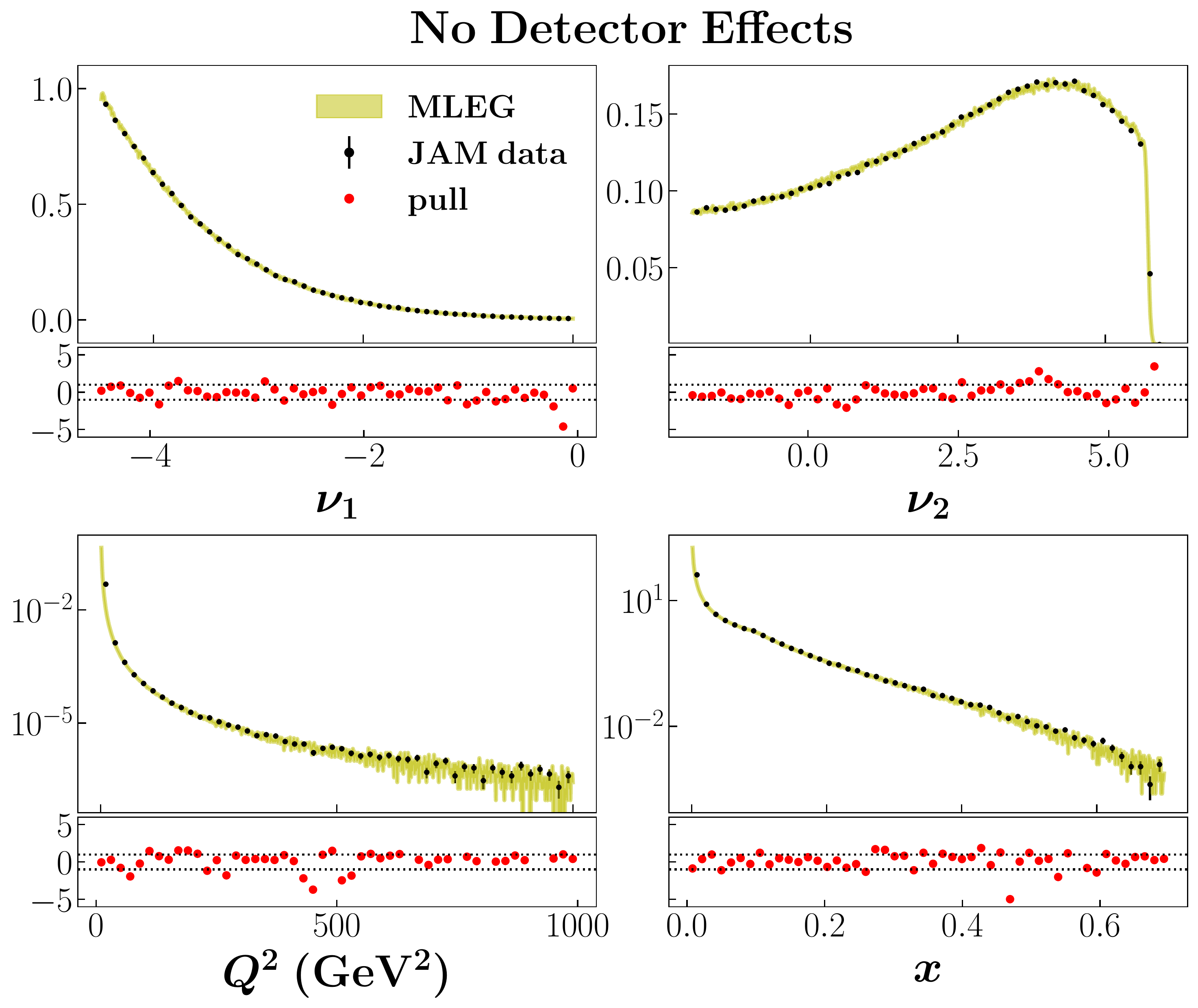}
    \caption{Comparison of distributions of training and derived variables from JAM training samples (black circles) and GAN-generated synthetic data (yellow bands) for the case of no detector effects; the band size reflects the uncertainty evaluated using the bootstrap procedure (see text).
    The bottom of each panel shows the pull distributions (red circles) defined in \eref{pull}, with the two horizontal dotted lines corresponding to $\pm 1 \sigma$.}
    \label{f.no_detector}
\end{figure}

As a first step in our numerical analysis, we train the MLEG using the DIS pseudodata samples without detector effects in order to establish the baseline agreement between training and synthetic data, without the complications introduced by the detector folding.
In \fref{no_detector} we compare the training and synthetic normalized inclusive $ep$ phase space distributions for the scattered electron in the variables $\nu_1$ and $\nu_2$.
The uncertainty bands were generated by training 10 independent GANs, where for each training the samples were prepared using the bootstrapping procedure ({\it i.e.}, taking random samples with replacement).
It is useful to define the ``pull'' metric between the training (JAM) and synthetic (GAN) data by
\begin{align}
{\rm pull}\
=\ \frac{ {\rm E}\big[ {\cal P}({\cal O}|{\rm bin}) \big]_{\rm GAN} 
        - {\rm E}\big[ {\cal P}({\cal O}|{\rm bin}) \big]_{\rm JAM} }
        { \sqrt{{\rm V}\big[{\cal P}({\cal O}|{\rm bin})\big]_{\rm GAN}
               +{\rm V}\big[{\cal P}({\cal O}|{\rm bin})\big]_{\rm JAM}} }\, ,
\label{e.pull}
\end{align}
where ${\rm E}[{\cal P}({\cal O}|{\rm bin})]$ and ${\rm V}[{\cal P}({\cal O}|{\rm bin})]$ are the expectation values and variances of the discrete probability density ${\cal P}$ of an observable ${\cal O}$.
As expected, the synthetic distributions for $\nu_1$ and $\nu_2$ match well with the distributions from the training samples, within the statistical uncertainties, since for these variables the deviation from the training set is explicitly disfavored by the discriminator. 
Also shown in \fref{no_detector} are distributions of derived quantities that are physically relevant for the DIS process, namely, the four-momentum transfer squared,
    $Q^2 = -(k-k')^2$,
and the Bjorken scaling variable
    $x = Q^2 / 2P\!\cdot\!(k-k')$.
While these observables are obtained by nonlinear transformations of the original variables $\nu_1$ and $\nu_2$, the result accurately reconstructs the matching, within uncertainties, with the corresponding spectra from the training data.

\begin{figure}[t] 
    \includegraphics[width=0.75\textwidth]{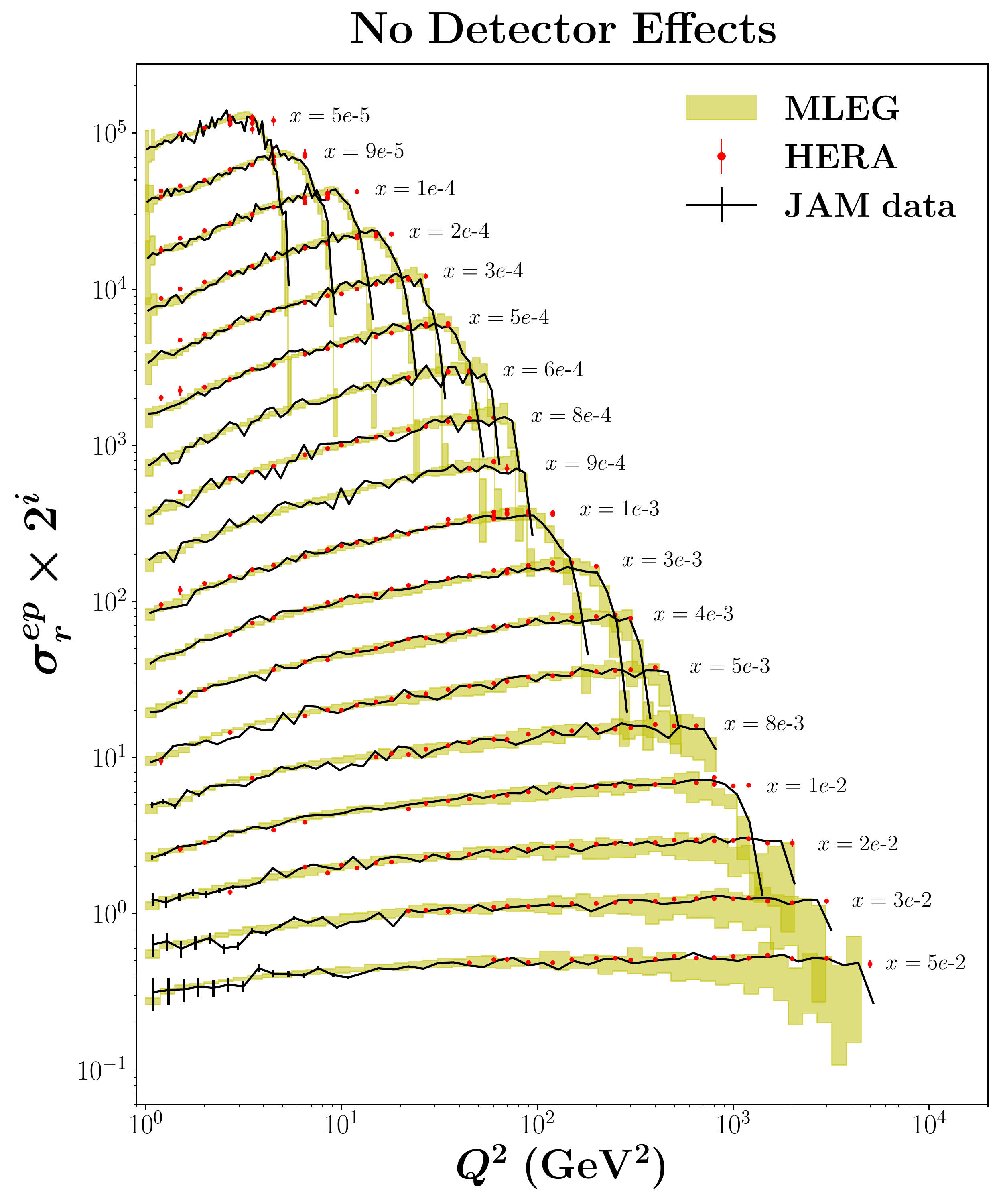}
    \caption{Comparison of the reduced inclusive $ep$ cross section $\sigma_r^{ep}$ versus $Q^2$ at fixed values of Bjorken-$x$ from the HERA collider \cite{Abramowicz:2015mha} (red circles) with data generated from the JAM global QCD analysis \cite{Cocuzza:2021rfn} (black solid lines) and the trained GAN (yellow bands). No detector effects are included, and for clarity the cross sections are scaled by a factor $2^i$, with $i$ ranging from $i=0$ for the highest-$x$ value to $i=17$ for the lowest-$x$ value.}
    \label{f.hera}
\end{figure}

In \fref{hera} we illustrate the reduced inclusive $ep$ DIS cross section, $\sigma_r^{ep}$ (in practice the reaction involved positrons scattering from protons), as a function of $Q^2$ in multiple bins of $x$ for the HERA data \cite{Abramowicz:2015mha} and for the parametrization of the data from the JAM global QCD analysis \cite{Cocuzza:2021rfn}.
These are compared with the reduced cross sections reconstructed by the GAN.
Within the statistical uncertainties, the empirical results are well reproduced by the MLEG simulation in most of the regions of the phase space.
Note that the agreement between the JAM fit and the HERA data deteriorates at the largest $Q^2$ values for each fixed-$x$ spectrum due to the vanishing of the phase space.
Nonetheless, as~\fref{hera} demonstrates, the GAN is able to reproduce this feature of the parametrization, indicating that the GAN has learned accurately the complex correlations present in the unpolarized DIS phase space.

\subsection{GAN training with detector effects}

\begin{figure}[t] 
    \includegraphics[width=0.7\textwidth]{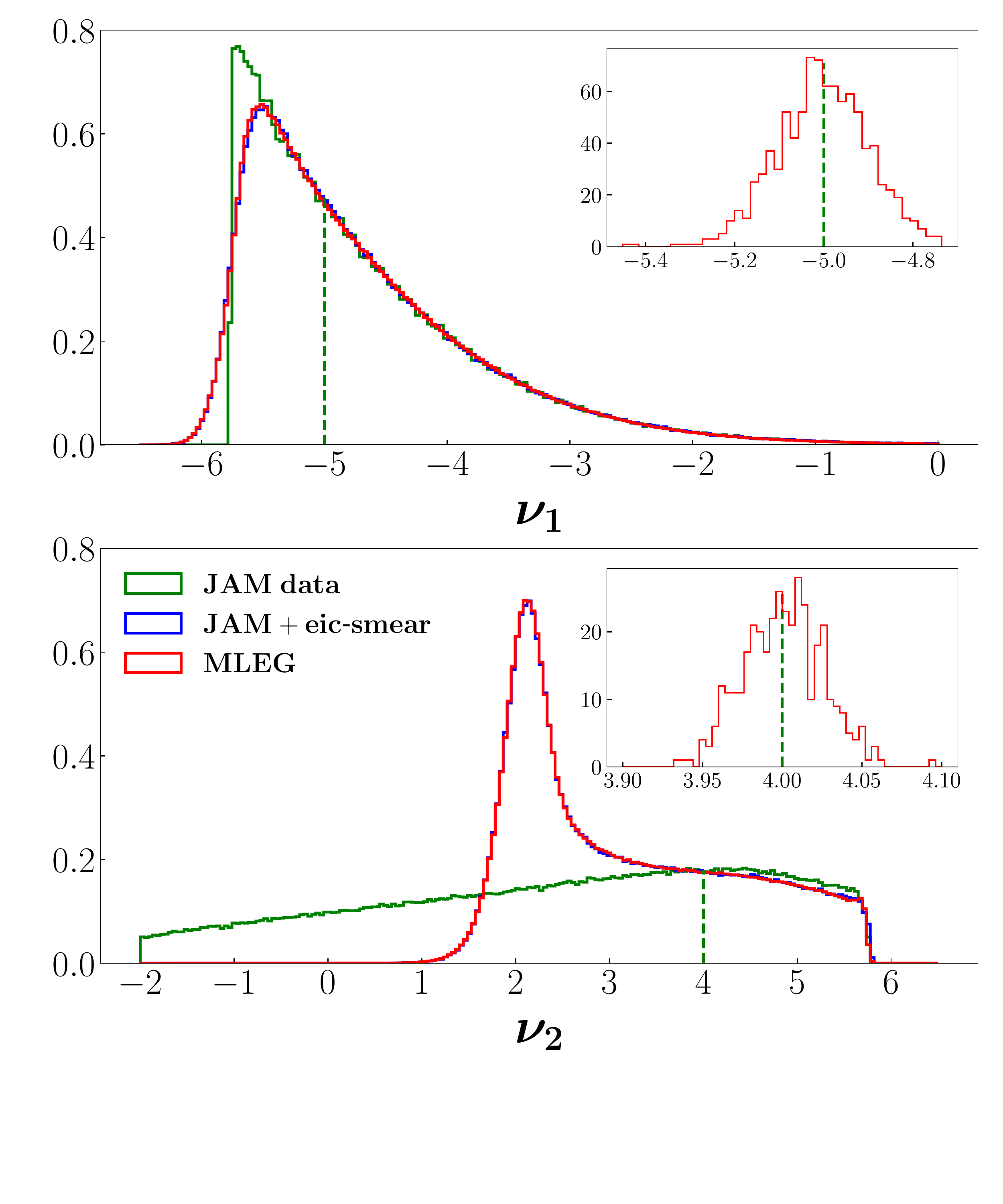}
    \vspace*{-1.5cm}
    \caption{Comparison of training features at the vertex level (generated, blue histograms) and detector level (smeared, green histograms) with the MLEG generated synthetic data (red histograms). The insets illustrate the local smearing effect at the points indicated by the green vertical dashed lines.}
    \label{f.pzsmear}
\end{figure}

Having established a baseline agreement for our MLEG framework, we proceed to include detector effects, as would be in actual experimental situations, which inevitably increases the complexity of the analysis.
As discussed above, we train separately an ML detector surrogate using a detector parametrization provided by the {\it eic-smear} software \cite{eicsoft}.
For the trial vertex-level event samples we use directly the samples from the JAM global QCD analysis instead of the flat phase space so as to optimize the GAN training.
However, we stress that in principle the model architecture for the detector surrogate can be trained with any samples.

In \fref{pzsmear} we show the vertex- and detector-level distributions for $\nu_1$ and $\nu_2$, where significant distortions are observed for the latter.
An issue regarding the change of variables in Eqs.~(\ref{e.nu12}) is that after smearing the detector-level $k'_z$ variable can exceed the physical limit given by the incident beam energy $E_e$, rendering the transformation singular for those unphysical cases.
However, since the change of variables, in particular for $\nu_2$, is solely designed to increase the detector awareness in the difficult regions, we can replace $E_e$ in Eqs.~(\ref{e.nu12}) by the maximum energy found for the detector-level samples to achieve the same goal, and avoid the singularity of the variable transform.
This disparity, however, creates an impression of higher levels of distortion in the $\nu_2$ variable compared to $\nu_1$.

We also illustrate the smearing effects by focusing on small intervals in $\nu_1$ and $\nu_2$, as shown in the Fig.~\ref{f.pzsmear} insets, to indicate the nontrivial distortion that is taking place across the phase space.
Included in Fig.~\ref{f.pzsmear} are the corresponding predictions from the detector-level GAN output, which shows very good agreement with the training samples.
Note that there are regions where GANs do not match precisely with {\it eic-smear}, namely, the tail regions at small and large $\nu_2$, which correspond to the edges of the reaction phase space. 
For the scope of this study, the GAN output represents a reasonable true detector proxy, allowing us to carry out the vertex-level learning closure test and validate the proof of principle of our MLEG framework.

\begin{figure}[h] 
    \includegraphics[width=0.8\textwidth]{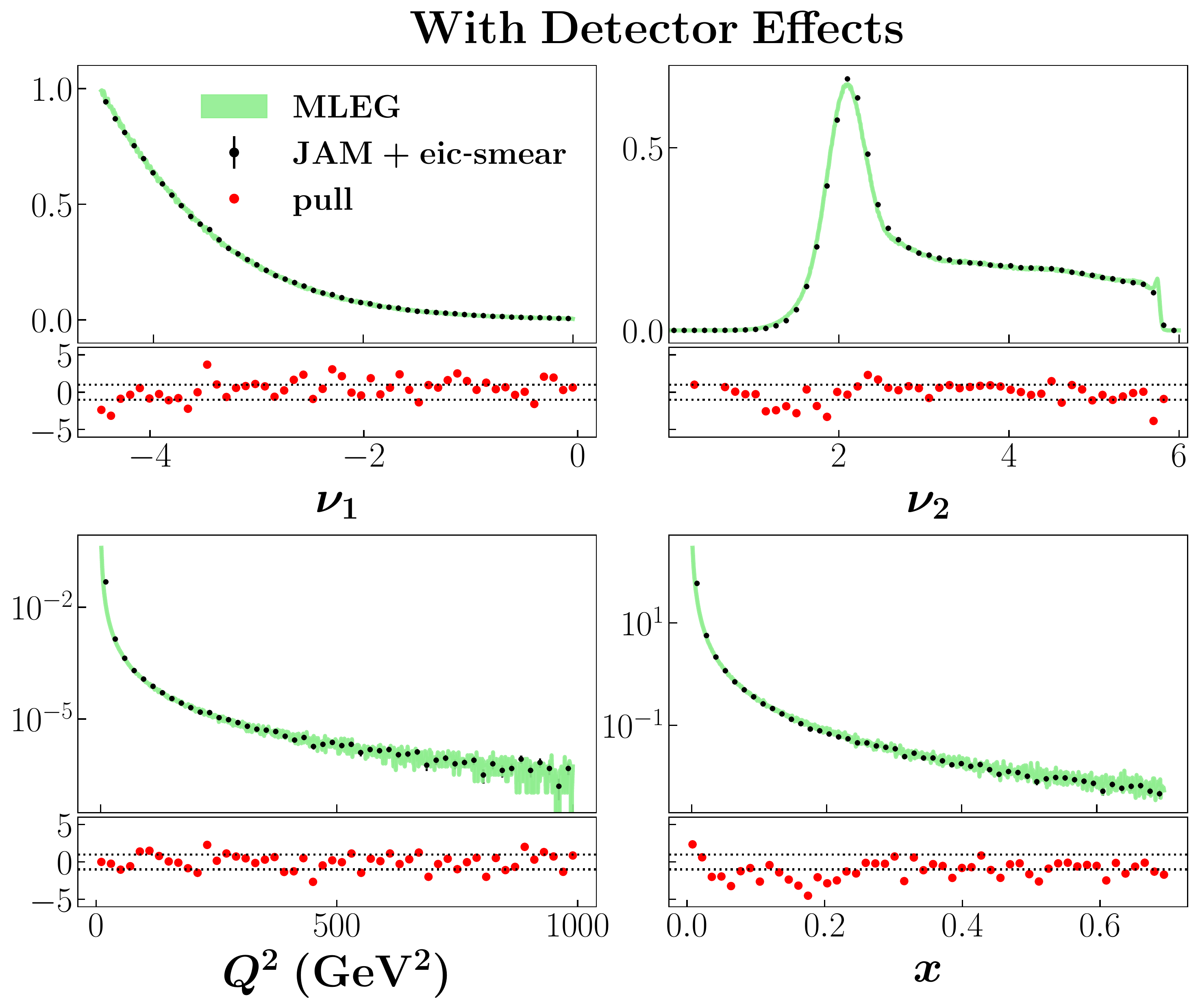}
    \caption{As in \fref{no_detector}, but with detector effects present.}
    \label{f.gan_detector}
\end{figure}
\begin{figure}[h] 
    \includegraphics[width=0.8\textwidth]{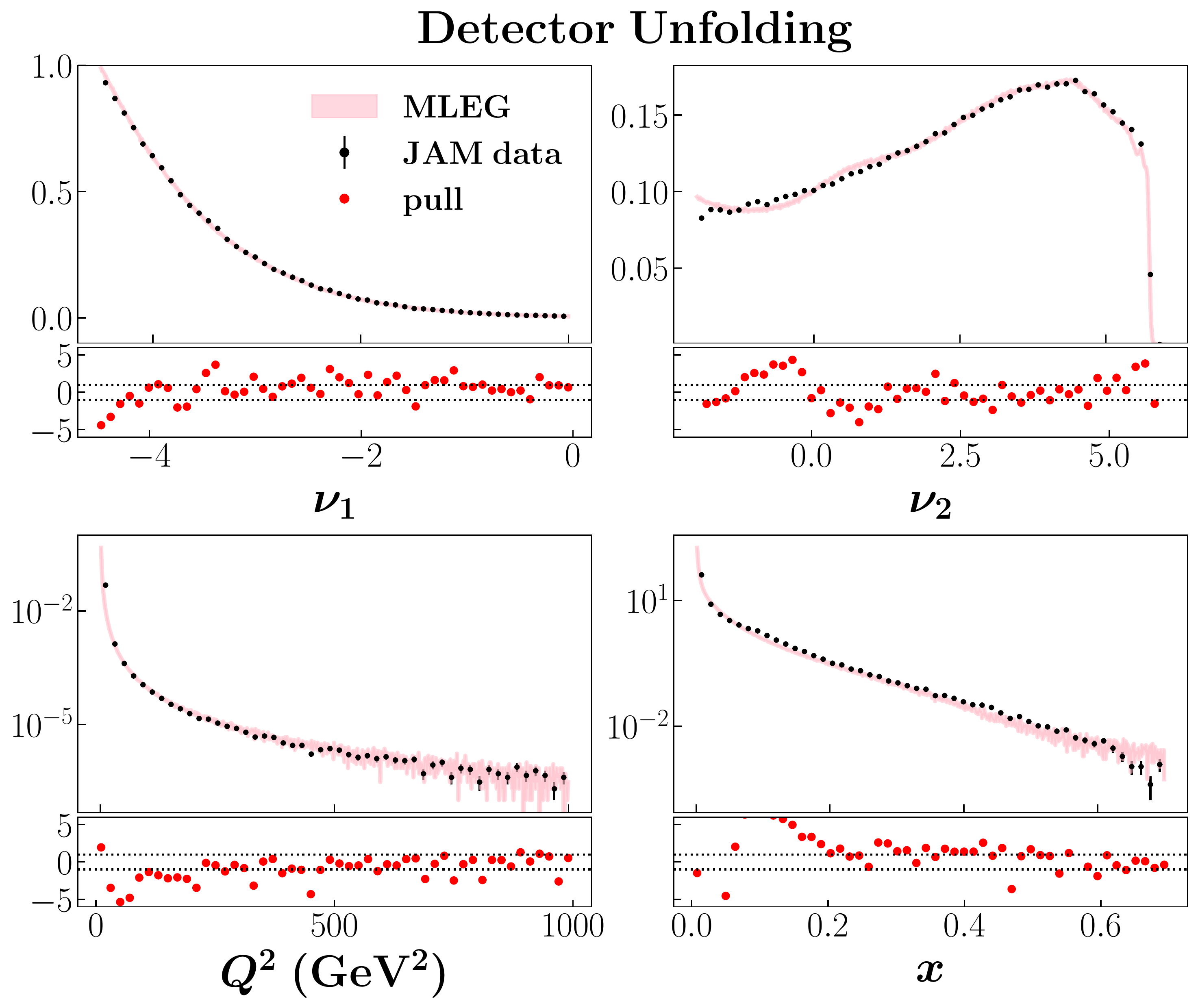}
    \caption{As in \fref{no_detector}, but with all the variables inferred by the unfolding procedure.}
    \label{f.gan_vertex}
\end{figure}

\begin{figure}[t] 
    \includegraphics[width=0.8\textwidth]{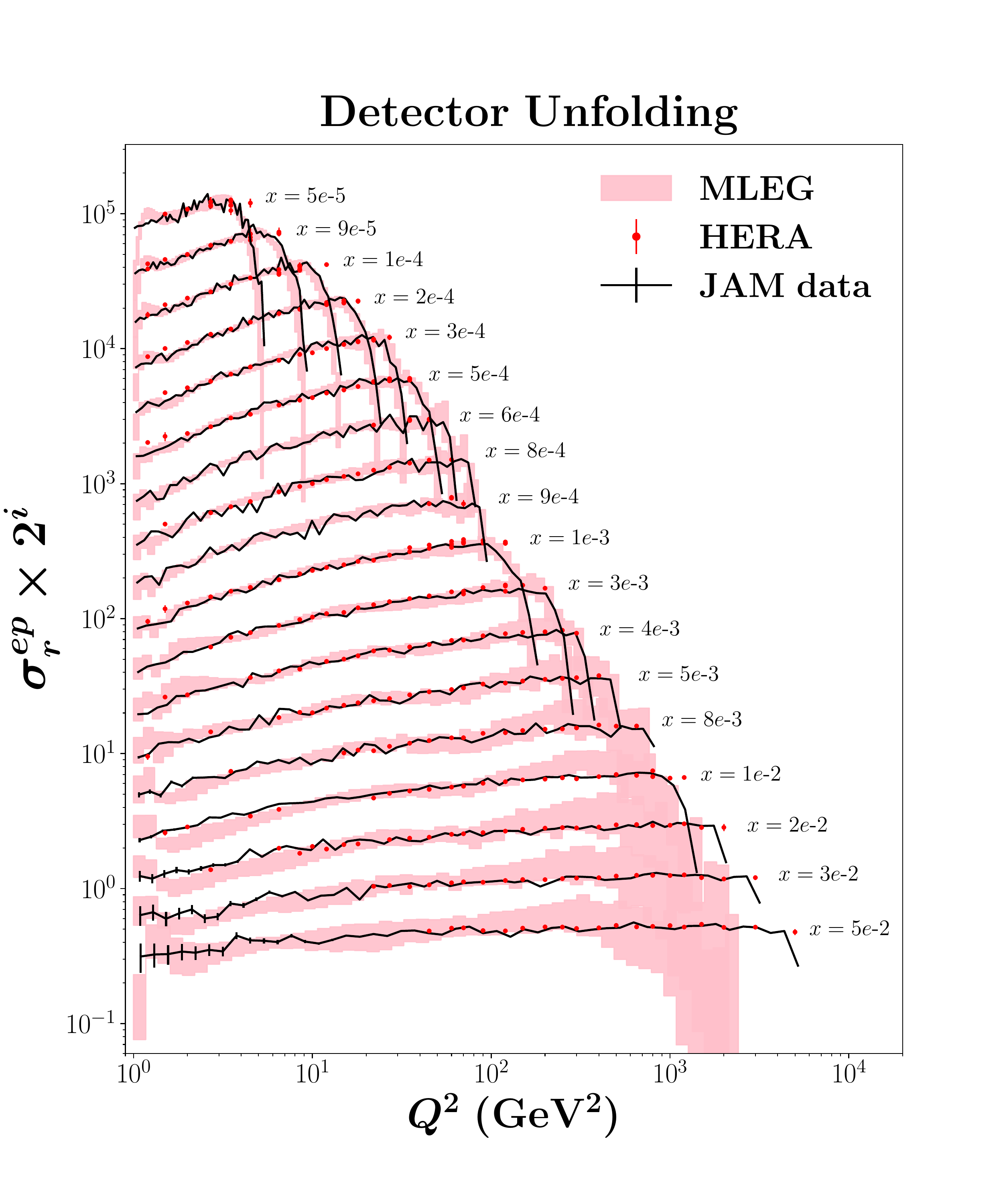}
    \vspace*{0cm}
    \caption{As in \fref{hera}, but with the synthetic reduced cross sections generated by the GAN including detector effects and unfolding.}
    \label{f.hera_smear}
\end{figure}

With the ML detector surrogate we proceed with training the MLEG with detector effects.
In \fref{gan_detector} we show similar results as in \fref{no_detector}, but this time with detector effects included.
As expected, the variables $\nu_1$ and $\nu_2$ are well reproduced, since the discriminator supervises on these variables during the training.
Similarly, the predicted DIS variables $x$ and $Q^2$ at the detector level are well reproduced within the uncertainties.

As the final step, we examine the quality of the MLEG at the vertex level by analysing the direct output of its generator, and plot in \fref{gan_vertex} the corresponding vertex-level distributions.
Relative to the detector level, the vertex-level distributions are observed to have, on average, larger values for the pull than those in \fref{gan_detector}.
This is expected since we do not directly supervise at the vertex level, but instead these are inferred quantities.
A more detailed examination of this is shown in \fref{hera_smear}, where we plot the reduced cross sections as in \fref{hera}, but in the presence of detector effects.
As expected, the uncertainties increase due to the detector effects.
However, within uncertainties, the synthetic reduced cross sections are in agreement with the true vertex level cross sections.
This can be seen as confirmation that our MLEG training passes the closure test in the presence of detector effects.

\clearpage
\section{Summary and Outlook}
\label{secV}

We have presented a new approach based on generative adversarial networks to extract physics observables from pseudodata in a physics agnostic manner.
To illustrate the strategy, we developed a GAN-based MLEG capable of generating synthetic data that mimic inclusive deep-inelastic $ep$ scattering pseudodata generated from PDFs in the kinematics of the ZEUS and H1 experiments at HERA.
To demonstrate the veracity of our approach we performed a closure test, extracting the original PDFs from synthetic particle four-momenta.

To simulate real experimental scenarios, we introduced distortions into the analysis that would be induced by a real detector, implementing a resolution smearing function, and after repeating the test obtained good agreement between original and extracted PDFs.
Pulls quantified the uncertainty associated with the unfolding procedure, showing not only that we were able to extract the desired physics observables, but providing an uncertainty quantification for the unfolding procedure.
To our knowledge this is the first time that detector effects were unfolded from pseudodata on an event basis.

While our long term goal remains to construct an MLEG for real experimental events across multiple channels for QCD studies, the present analysis is a necessary and important proof of concept that demonstrates the viability of applying ML techniques to mitigate theoretical bias in experimental data analysis.
The promising results found with the case study of inclusive $ep$ DIS suggests potentially important applications of the GAN-based MLEGs to physical processes beyond inclusive reactions.

As obvious improvements, and in view of its application to data analysis, we envision the implementation of a more realistic detector simulator based on GEANT to further study this technology. 
We expect that the use of our framework in $ep$ scattering will be a valuable complementary tool for nuclear and particle physics programs at current and planned facilities, such as Jefferson Lab~\cite{Burkert:2018nvj} and the  Electron-Ion Collider~\cite{Accardi:2012qut}.

\clearpage
\section*{Acknowledgements}

We thank J.~Qiu for helpful discussions.
This work was supported by the Jefferson Lab LDRD project No.~LD2122, and by the U.S. Department of Energy contract DE-AC05-06OR23177, under which Jefferson Science Associates, LLC operates Jefferson Lab, and by Deutsche Forschungsgemeinschaft (DFG) through the Research Unit FOR2926 (project number 40824754), and in part by NSF grant no PHY-2012865.


\end{document}